\documentclass[graybox]{svmult}

\usepackage{multirow}
\usepackage{mathptmx}       
\usepackage{helvet}         
\usepackage{courier}        
\usepackage{type1cm}        
\usepackage{makeidx}         
\usepackage{graphicx}        
\usepackage{multicol}        
\usepackage[bottom]{footmisc}

\usepackage{mathptmx}       
\usepackage{helvet}         
\usepackage{courier}        
\usepackage{type1cm}        
\usepackage{makeidx}         
\usepackage{graphicx}        
\usepackage{multicol}        
\usepackage[bottom]{footmisc}
\usepackage{url}

\usepackage{ifthen}
\usepackage{makeglos}

\graphicspath{{./}{./graphics/}{./graphics/eps/}{./graphics/pdf/}}

\usepackage{graphicx}
\usepackage{amsmath}
\usepackage{amssymb}
\usepackage{caption}
\usepackage{dblfloatfix}
\usepackage{bigstrut}
\usepackage{multirow}
\usepackage{enumitem}
\usepackage{balance}

\newcommand{\buildbook}{false}

\makeindex             

\makeglossary

\begin{document}

%
%
%

\ifthenelse{\equal{true}{\buildbook}}{
\title{Introduction to Voice Presentation Attack Detection and Recent Advances}
}
{
\title*{Introduction to Voice Presentation Attack Detection and Recent Advances}
}


\author{Md Sahidullah, H\'{e}ctor Delgado, Massimiliano Todisco, Tomi Kinnunen, Nicholas Evans, Junichi Yamagishi and Kong-Aik Lee}


\institute{Md Sahidullah \at School of Computing, University of Eastern Finland (Finland), \email{sahid@cs.uef.fi} [Currently with Inria, France.]
\and H\'{e}ctor Delgado \at Department of Digital Security, EURECOM (France) \email{hector.delgado@eurecom.fr}
\and Massimiliano Todisco \at Department of Digital Security, EURECOM (France) \email{massimiliano.todisco@eurecom.fr}
\and Tomi Kinnunen \at School of Computing, University of Eastern Finland (Finland), \email{tkinnu@cs.uef.fi}
\and Nicholas Evans \at Department of Digital Security, EURECOM (France) \email{evans@eurecom.fr}
\and Junichi Yamagishi \at  National Institute of Informatics (Japan) and University of Edinburgh (United Kingdom) \email{jyamagis@nii.ac.jp}
\and Kong-Aik Lee \at  Data Science Research Laboratories, NEC Corporation (Japan) \email{k-lee@ax.jp.nec.com}
}

%

%
\maketitle

%


\abstract*{
Over the past few years significant progress has been made in the field of presentation attack detection (PAD) for automatic speaker recognition (ASV).
This includes the development of new speech corpora,
standard evaluation protocols and advancements in front-end feature extraction and
back-end classifiers. The use of standard databases and evaluation protocols has enabled for
the first time the meaningful benchmarking of different PAD solutions. This chapter summarises the progress, with a focus on studies completed in the last three years. The article presents a summary of findings and lessons learned from two ASVspoof challenges, the first community-led benchmarking efforts. These show that ASV PAD remains an unsolved problem and that further attention is required to develop generalised
PAD solutions which have potential to detect diverse and previously unseen spoofing
attacks.}

\abstract{Over the past few years significant progress has been made in the field of presentation attack detection (PAD) for automatic speaker recognition (ASV).
This includes the development of new speech corpora,
standard evaluation protocols and advancements in front-end feature extraction and
back-end classifiers. The use of standard databases and evaluation protocols has enabled for
the first time the meaningful benchmarking of different PAD solutions. This chapter summarises the progress, with a focus on studies completed in the last three years. The article presents a summary of findings and lessons learned from two ASVspoof challenges, the first community-led benchmarking efforts. These show that ASV PAD remains an unsolved problem and that further attention is required to develop generalised
PAD solutions which have potential to detect diverse and previously unseen spoofing
attacks.}

\section{Introduction}
Automatic speaker verification (ASV) technology aims to recognise individuals using samples of the human voice signal~\cite{Kinnunen201012,hansen2015speaker}. Most ASV systems operate on estimates of the spectral characteristics of voice in order to recognise individual speakers.  ASV technology has matured in recent years and now finds application in a growing variety of real-world authentication scenarios involving both \emph{logical} and \emph{physical} access.  In \index{logical access} scenarios, ASV technology can be used for remote person authentication via the Internet or traditional telephony.  In many cases, ASV serves as a convenient and efficient alternative to more conventional password-based solutions, one prevalent example being person authentication for Internet and mobile banking.
\index{Physical access} scenarios include the use of ASV to protect personal or secure/sensitive facilities, such as domestic and office environments.
With the growing, widespread adoption of smartphones and voice-enabled smart devices, such as intelligent personal assistants all equipped with at least one microphone, ASV technology stands to become even more ubiquitous in the future.

\begin{figure*}[!t]
\centerline{\includegraphics[width=\linewidth]
{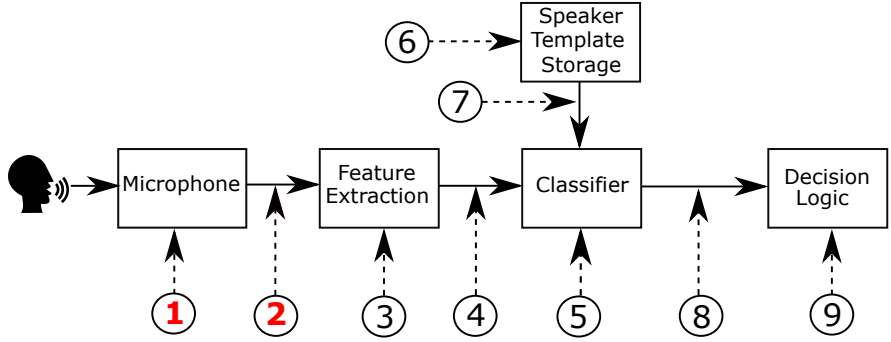}}
\caption{Possible attack locations in a typical ASV system. 1: microphone point, 2: transmission point, 3: override feature extractor, 4: modify probe to features, 5: override classifier, 6: modify speaker database, 7: modify biometric reference, 8: modify score and 9: override decision.}
\label{fig:attackpoints}
\end{figure*}

Despite its appeal, the now-well-recognised vulnerability to manipulation through presentation attacks (PAs), also known as spoofing, has dented confidence in ASV technology. As identified in ISO/IEC 30107-1 standard~\cite{iso2016presentation}, the possible locations of presentation attack points in a typical ASV system are illustrated in Fig.~\ref{fig:attackpoints}. Two of the most vulnerable places in an ASV system are marked by 1 and 2, corresponding to physical access and logical access. This work is related to these two types of attacks.

Unfortunately, ASV is arguably more prone to PAs than other biometric systems based on traits or characteristics that are less-easily acquired; samples of a given person's voice can be collected readily by fraudsters through face-to-face or telephone conversations and then replayed in order to manipulate an ASV system.  Replay attacks are furthermore only one example of ASV PAs.
More advanced voice conversion or speech synthesis algorithms can be used to generate particularly effective PAs using only modest amounts of voice data collected from a target person.

There are a number of ways to prevent PA problems. The first one is based on a text-prompted system which uses an utterance verification process~\cite{kinnunen2016utterance}. The user needs to utter a specific text, prompted for authentication by the system which requires a text-verification system. Secondly, as human can never
reproduce an identical speech signal, some countermeasures use template matching or audio fingerprinting to verify whether
the speech utterance was presented to the system earlier~\cite{shang2010score}. Thirdly, some work looks into statistical acoustic characterisation of authentic speech and speech created with presentation attack methods or spoofing techniques~\cite{Wu2015survey}. Our focus is on the last category, which is more convenient in a practical scenario for both text-dependent and text-independent ASV. In this case, given a speech signal, $S$, PA detection – here, the determination of whether $S$ is a natural or PA speech – can be formulated as a hypothesis test:

\begin{itemize}
  \item $H_0$: $S$ is natural speech.
  \item $H_1$: $S$ is created with PA methods.
\end{itemize}

A \index{likelihood ratio test} can be applied to decide between $H_0$ and $H_1$. Suppose that $\mathbf{X} = \{\mathbf{x}_1, \mathbf{x}_2,...,\mathbf{x}_N\}$ are the acoustic feature
vectors of $N$ speech frames extracted from $S$, then the logarithmic likelihood ratio score is given by,

\begin{equation}\label{eq:llr}
\Lambda(\mathbf{X})=\log p(\mathbf{X}|\lambda_{H_0})-\log p(\mathbf{X}|\lambda_{H_1})
\end{equation}

\begin{figure*}[!t]
\centerline{\includegraphics[width=\linewidth]
{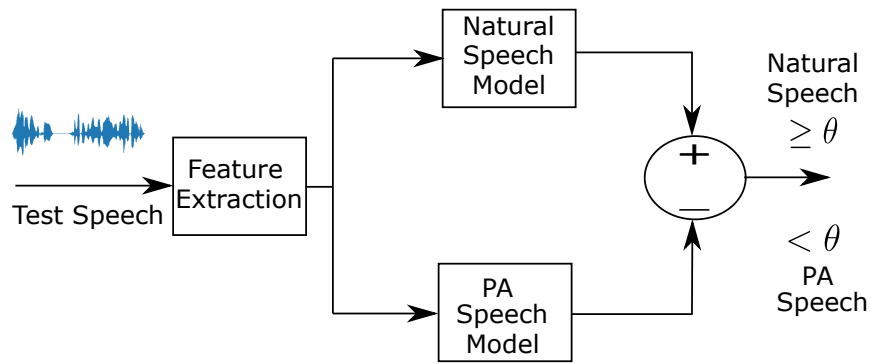}}
\caption{Block diagram of a typical presentation attack detection system.}
\label{fig:blockdiagram}
\end{figure*}

In\ref{eq:llr}, $\lambda_{H_0}$ and $\lambda_{H_1}$ are the acoustic models to characterise the hypotheses correspondingly for natural speech and PA speech. The parameters of these models are estimated using training data for natural and PA speech. A typical PAD system is shown in Fig.~\ref{fig:blockdiagram}. A test speech can be accepted as natural or rejected as PA speech with help of a threshold, $\theta$ computed on some development data. If the score is greater than or equal to the threshold, it is accepted; otherwise, rejected. The performance of the PA system is assessed by computing the \index{\emph{equal error rate}} (EER) metric. This is the error rate for a specific value of a threshold where two error rates, i.e., the probability of a PA speech detected as being natural speech (known as false acceptance rate or FAR) and the probability of a natural speech speech being misclassified as a PA speech (known as false rejection rate or FRR), are equal. Sometimes \index{\emph{half total error rate}} (HTER) is also computed~\cite{Korshunov2016Overview}. This is the average of FAR and FRR which are computed using a decision threshold obtained with the help of the development data.

Awareness and acceptance of the vulnerability to PAs have generated a growing interest in develop solutions to presentation attack detection (PAD), also referred to as spoofing countermeasures. These are typically dedicated auxiliary systems which function in tandem to ASV in order to detect and deflect PAs.
The research in this direction has progressed rapidly in the last three years, due partly to the release of several public speech corpora and the organisation of PAD challenges for ASV. This article, a continuation of the chapter~\cite{handbook2014} in the first edition of the Handbook for Biometrics~\cite{handbook2014book} presents an up-to-date review of the different forms of voice presentation attacks, broadly classified in terms of impersonation, replay, speech synthesis and voice conversion.  The primary focus is nonetheless on the progress in PAD.
The chapter reviews the most recent work involving a variety of different features and classifiers. Most of the work covered in the chapter relates to that conducted using the two most popular and publicly available databases, which were used for the two ASVspoof challenges co-organized by the authors. The chapter concludes with a discussion of research challenges and future directions in PAD for~ASV.

\section{Basics of ASV spoofing and countermeasures}
Spoofing or presentation attacks are performed on a biometric system at the sensor or acquisition level
to bias score distributions toward those of genuine clients, thus provoking increases
in the false acceptance rate (FAR). This section reviews four well-known ASV spoofing techniques and their respective countermeasures: impersonation,
replay, speech synthesis and voice conversion. Here, we mostly review the work in the pre-ASVspoof period, as well as some very recent studies on presentation attacks.

\subsection{Impersonation}
In speech \index{impersonation} or mimicry attacks, an intruder speaker intentionally modifies his or her speech to sound like the target speaker. Impersonators are likely to copy lexical, prosodic, and idiosyncratic behaviour of their target speakers presenting a potential point of vulnerability concerning speaker recognition systems.

\subsubsection{Spoofing}
There are several studies about the consequences of mimicry on ASV. Some studies concern attention to the voice modifications performed by professional impersonators. It has been reported that impersonators are often particularly able to adapt the fundamental frequency (F0) and occasionally also the formant frequencies towards those of the target speakers~\cite{farrus2010automatic,perrot2007voice,zetterholm2007detection}. In studies, the focus has been on analysing the vulnerability of speaker verification systems in the presence of voice mimicry. The studies by Lau et al.~\cite{lau2004vulnerability,lau2005testing} suggest that if the target of impersonation is known in advance and his or her voice is ``similar'' to the impersonator's voice (in the sense of automatic speaker recognition score), then the chance of spoofing an automatic recognizer is increased. In~\cite{mariethoz2005can}, the experiments indicated that professional impersonators are potentially better impostors than amateur or naive ones. Nevertheless, the voice impersonation was not able to spoof the ASV system. In~\cite{farrus2010automatic}, the authors attempted to quantify how much a speaker is able to approximate other speakers' voices by selecting a set of prosodic and voice source features. Their prosodic and acoustic based ASV results showed that two professional impersonators imitating known politicians increased the identification error rates.

More recently, a fundamentally different study was carried out by Panjwani et al.~\cite{panjwani2014crowdsourcing} using crowdsourcing to recruit both amateur and more professional impersonators. The results showed that impersonators succeed in increasing their average score, but not in exceeding the target speaker score. All of the above studies analysed the effects of speech impersonation either at the acoustic or speaker recognition score level, but none proposed any countermeasures against impersonation. In a recent study~\cite{hautamaki2015automatic}, the experiments aimed to evaluate the vulnerability of three modern speaker verification systems against impersonation attacks and to further compare these results to the performance of non-expert human listeners. It is observed that, on average, the mimicry attacks lead to increased error rates. The increase in error rates depends on the impersonator and the ASV system.

The main challenge, however, is that no large speech corpora of impersonated speech exists for the quantitative study of impersonation effects on the same scale as for other attacks, such as text-to-speech synthesis and voice conversion, where generation of simulated spoofing attacks as well as developing appropriate countermeasures is more convenient.

\subsubsection{Countermeasures}
While the threat of impersonation is not fully understood due to limited studies
involving small datasets, it is perhaps not surprising that there is no prior work investigating countermeasures against impersonation. If the threat is proven to be genuine,
then the design of appropriate countermeasures might be challenging. Unlike
the spoofing attacks discussed below, all of which can be assumed to leave traces of
the physical properties of the recording and playback devices, or signal processing
artefacts from synthesis or conversion systems, impersonators are live human beings
who produce entirely natural speech.

\subsection{Replay}
\index{Replay} attacks refer to the use of pre-recorded speech from a target speaker, which is then replayed through some playback device to feed the system microphone. These attacks require no specific expertise nor sophisticated equipment, thus they are easy to implement. Replay is a relatively low-technology attack
within the grasp of any potential attacker even without specialised knowledge in speech processing. Several works in the earlier literature report significant increases in error rates when using replayed speech. Even if replay attacks may present a genuine risk to ASV systems, the use of prompted-phrase has the potential to mitigate the impact.

\subsubsection{Spoofing}
The study on the impact of replay attack on ASV performance was very limited until recently before the release of AVspoof~\cite{KucurErgunay_IEEEBTAS_2015} and ASVspoof 2017 corpus. The earlier studies were conducted either on simulated or on real replay recording from far-field.

The vulnerability of ASV systems to replay attacks was first investigated in a text-dependent scenario~\cite{lindberg1999vulnerability}, where the concatenation of recorded digits was tested against a hidden Markov model (HMM) based ASV system. Results showed an
increase in the FAR from 1 to 89\% for male speakers and from 5
to 100\% for female speakers.

The work in~\cite{villalba2010speaker} investigated text-independent ASV vulnerabilities through the
replaying of far-field recorded speech in a mobile telephony scenario where signals
were transmitted by analogue and digital telephone channels. Using a baseline ASV
system based on \emph{joint factor analysis} (JFA), the work showed an increase in the EER of 1\% to almost
70\% when impostor accesses were replaced by replayed spoof attacks.

A physical access scenario was considered in~\cite{Wang2011}. While the baseline performance of the Gaussian mixture model- universal background model (GMM-UBM) ASV system was not reported, experiments showed that replay attacks
produced a FAR of 93\%.

The work in~\cite{KucurErgunay_IEEEBTAS_2015} introduced audio-visual spoofing (AVspoof) database for replay attack detection where the replayed signals are collected and played back using different low-quality (phones and laptop) and high-quality (laptop with loud speakers) devices. The study reported that FARs for replayed speech was 77.4\% and 69.4\% for male and female, respectively, using a total variability system speaker recognition system. In this study, the EER for bona fide trials was 6.9\% and 17.5\% for those conditions. This study also includes presentation attack where speech signals created with voice conversion and speech synthesis were used in playback attack. In that case, higher FAR was observed, particularly when high-quality device is used for playback.

\subsubsection{Countermeasures}
A countermeasure for replay attack detection in the case of text-dependent ASV was
reported in~\cite{shang2010score}. The approach is based upon the comparison of new access samples
with stored instances of past accesses. New accesses which are deemed too similar to
previous access attempts are identified as replay attacks. A large number of different
experiments, all relating to a telephony scenario, showed that the countermeasures
succeeded in lowering the EER in most of the experiments performed.
While some form of text-dependent or challenge-response countermeasure is usually
used to prevent replay attacks, text-independent solutions have also been investigated.
The same authors in~\cite{villalba2010speaker} showed that it is possible to detect replay attacks
by measuring the channel differences caused by far-field recording~\cite{villalba2011preventing}. While they
show spoof detection error rates of less than 10\% it is feasible that today's state-of-the-art
approaches to channel compensation will render some ASV systems still
vulnerable.

Two different replay attack countermeasures are compared in~\cite{Wang2011}. Both are based
on the detection of differences in channel characteristics expected between licit and
spoofed access attempts. Replay attacks incur channel noise from both the recording
device and the loudspeaker used for replay and thus the detection of channel effects
beyond those introduced by the recording device of the ASV system thus serves as
an indicator of replay. The performance of a baseline GMM-UBM system with an
EER of 40\% under spoofing attack falls to 29\% with the first countermeasure and a
more respectable EER of 10\% with the second countermeasure.

In another study~\cite{galka2015playback}, a speech database of 175 subjects has been collected for different kinds of replay attack. Other than the use of genuine voice samples for the legitimate speakers in playback, the voice samples recorded over the telephone channel were also used for unauthorised access. Further, a far-field microphone is used to collect the voice samples as eavesdropped (covert) recording. The authors proposed an algorithm motivated from music recognition system used for comparing recordings on the basis
of the similarity of the local configuration of maxima pairs extracted from spectrograms of verified and reference
recordings. The experimental results show the EER of playback attack detection to be as low as 1.0\% on the collected data.

\subsection{Speech synthesis}

\index{Speech synthesis}, commonly referred to as text-to-speech (TTS), is a technique for generating intelligible, natural sounding artificial speech for any arbitrary text. Speech synthesis is used widely in various applications including in-car navigation systems, e-book readers, voice-over for the visually impaired and communication aids for the speech impaired. More recent applications include spoken dialogue systems, communicative robots, singing speech synthesisers and speech-to-speech translation systems.

Typical speech synthesis systems have two main components~\cite{Taylor:2009}: text analysis followed by speech waveform generation, which are sometimes referred to as the front-end and back-end respectively. In the text analysis component, input text is converted into a linguistic specification consisting of elements such as phonemes. In the speech waveform generation component, speech waveforms are generated from the produced linguistic specification. There are emerging end-to-end frameworks that generate speech waveforms directly from text inputs without using any additional modules.

Many approaches have been investigated, but there have been major paradigm shifts every ten years. In the early 1970s, the speech waveform generation component used very low dimensional acoustic parameters for each phoneme, such as formants, corresponding to vocal tract resonances with hand-crafted acoustic rules~\cite{Klatt}. In the 1980s, the speech waveform generation component used a small database of phoneme units called \emph{diphones} (the second half of one phoneme plus the first half of the following) and concatenated them according to the given phoneme sequence by applying signal processing, such as linear predictive (LP) analysis, to the units~\cite{Moulines_1990}. In the 1990s, larger speech databases were collected and used to select more appropriate speech units that matched both phonemes and other linguistic contexts such as lexical stress and pitch accent in order to generate high-quality natural sounding synthetic speech with the appropriate prosody.  This approach is generally referred to as \emph{unit selection}, and is nowadays used in many speech synthesis systems~\cite{hunt_icassp1996,Breen_1998,donovan_HMMSYN_ICSLP,NextGen_TTS,Coorman_2000}.

In the late 2000s, several machine learning based data-driven approaches emerged.  `Statistical parametric speech synthesis' was one of the more popular machine learning approaches~\cite{yoshimura_SIMMOD_EURO,USTC_Blizzard06,black06,Zen_Nitech-HTS2005}.  In this approach, several acoustic parameters are modelled using a time-series stochastic generative model, typically a HMM. HMMs represent not only the phoneme sequences but also various contexts of the linguistic specification.  Acoustic parameters generated from HMMs and selected according to the linguistic specification are then used to drive a vocoder, a simplified speech production model in which speech is represented by vocal tract parameters and excitation parameters in order to generate a speech waveform. HMM-based speech synthesisers~\cite{Zen2009,ref:CSMAPLR-IEEE2007} can also learn speech models from relatively small amounts of speaker-specific data by adapting background models derived from other speakers based on the standard model adaptation techniques drawn from speech recognition, i.e., maximum likelihood linear regression (MLLR)~\cite{Leggetter_CSL, Woodland_Review_Adapt}.

In the 2010s, deep learning has significantly improved the performance of speech synthesis and led to a  significant breakthrough. First, various types of deep neural networks are used to improve the prediction accuracy of the acoustic parameters~\cite{6639215,6542729}. Investigated architectures include recurrent neural network~\cite{FanQXS14, 7178816, 7472657}, residual/highway network~\cite{wanginvestigating,WANG20181}, autoregressive network~\cite{7953087,Wang2017}, and generative adversarial networks (GAN)~\cite{saito2017training,8063435,kaneko2017generative}. Furthermore, in the late 2010s conventional waveform generation modules that typically used signal processing and text analysis modules that used natural language processing were substituted by neural networks. This allows for neural networks capable of directly outputting the desired speech waveform samples from the desired text inputs. Successful architectures for direct waveform modelling include dilated convolutional autoregressive neural network, known as ``Wavenet''~\cite{van2016wavenet} and hierarichical recurrent neural network, called ``SampleRNN''~\cite{mehri2016samplernn}.  Finally, we have also seen successful architectures that totally remove the hand-crafted linguistic features obtained through text analysis by relying in sequence-to-sequence systems. This system is called Tacotron~\cite{Wang2017Taco}. As  expected, the combination of these advanced models results in a very high-quality end-to-end TTS synthesis system~\cite{gibiansky2017deep,46619} and recent results reveal that the generated synthetic speech sounds as natural as human speech~\cite{46619}.

For more details and technical comparisons, please see the results of Blizzard Challenge, which annually compares the performance of speech synthesis systems built on the common database over decades~\cite{king2014measuring, blizzard2017}.

\subsubsection{Spoofing}

There is a considerable volume of research in the literature which has demonstrated the vulnerability of ASV to synthetic voices generated with a variety of approaches to speech synthesis.  Experiments using formant, diphone, and unit-selection based synthetic speech in addition to the simple cut-and-paste of speech waveforms have been reported~\cite{lindberg1999vulnerability,5444499,villalba2010speaker}.

ASV vulnerabilities to HMM-based synthetic speech were first demonstrated over a decade ago~\cite{masuko99} using an HMM-based, text-prompted ASV system~\cite{matsui95} and an HMM-based synthesiser where acoustic models were adapted to specific human speakers~\cite{masuko96, masuko97}.  The ASV system scored feature vectors against speaker and background models composed of concatenated phoneme models.  When tested with human speech, the ASV system achieved a FAR of 0\% and a false rejection rate (FRR) of 7\%.  When subjected to spoofing attacks with synthetic speech, the FAR increased to over 70\%, however, this work involved only 20 speakers.

Larger scale experiments using the Wall Street Journal corpus containing in the order of 300 speakers and two different ASV systems (GMM-UBM and SVM using Gaussian supervectors) was reported in~\cite{6205335}.  Using an HMM-based speech synthesiser, the FAR was shown to rise to 86\% and 81\% for the GMM-UBM and SVM systems respectively representing a genuine threat to ASV. Spoofing experiments using HMM-based synthetic speech against a forensics speaker verification tool~\textit{BATVOX} was also reported in~\cite{galou} with similar findings. Therefore, the above speech synthesisers were chosen as one of spoofing methods in the ASVspoof 2015 database.

Spoofing experiments using the above advanced DNNs or using spoofing-specific strategies such as GAN have not yet been properly investigated. Only a relatively small-scale spoofing experiment against a speaker recognition system using Wavenet, SampleRNN and GAN is reported in ~\cite{cai2018attacking}.

\subsubsection{Countermeasures}

Only a small number of attempts to discriminate synthetic speech from natural speech had been investigated before the ASVspoof challenge started. Previous work has demonstrated the successful detection of synthetic speech based on prior knowledge of the acoustic differences of specific speech synthesizers, such as the dynamic ranges of spectral parameters at the utterance level~\cite{satoh01} and variance of higher order parts of mel-cepstral coefficients~\cite{5684887}.

There are some attempts which focus on acoustic differences between vocoders and natural speech.  Since the human auditory system is known to be relatively insensitive to phase~\cite{quatieri02}, vocoders are typically based on a minimum-phase vocal tract model. This simplification leads to differences in the phase spectra between human and synthetic speech, differences which can be utilised for discrimination~\cite{6205335,wu2012detecting}.

Based on the difficulty in reliable prosody modelling in both unit selection and statistical parametric speech synthesis, other approaches to synthetic speech detection use F0 statistics~\cite{OGIHARAAkio:2005-01-01,PhillipIS2012}.  F0 patterns generated for the statistical parametric speech synthesis approach tend to be over-smoothed and the unit selection approach frequently exhibits `F0 jumps' at concatenation points of speech units.

After the ASVspoof challenges took place, various types of countermeasures that work for both speech synthesis and voice conversion have been proposed. Please read the next section for the details of the recently developed countermeasures.

\subsection{Voice conversion}

\index{Voice conversion}, in short, VC \glossary{VC: voice conversion}, is a spoofing attack against automatic speaker verification using an attackers natural voice which is converted towards that of the target.
It aims to convert one speaker's voice towards that of another and is a sub-domain of voice transformation~\cite{stylianou2009voice}.
Unlike TTS, which requires text input, voice conversion operates directly on speech inputs. However, speech waveform generation modules such as vocoders, may be the same as or similar to those for TTS.

A major application of VC is to personalise and create new voices for TTS synthesis systems and spoken dialogue systems. Other applications include speaking aid devices that generate more intelligible voice sounds to help people with speech disorders, movie dubbing, language learning, and singing voice conversion. The field has also attracted increasing interest in the context of ASV vulnerabilities for almost two decades~\cite{pellom1999experimental}.

Most voice conversion approaches require a parallel corpus where source and target speakers read out identical utterances and adopt a training phase which typically requires frame- or phone-aligned audio pairs of the source and target utterances and estimates transformation functions that convert acoustic parameters of the source speaker to those of the target speaker. This is called ``parallel voice conversion''.  Frame alignment is traditionally achieved using dynamic time warping (DTW) on the source-target training audio files. Phone alignment is traditionally achieved using \emph{automatic speech recognition} (ASR) and phone-level forth alignment. The estimated conversion function is then applied to any new audio files uttered by the source speaker~\cite{MOHAMMADI201765}.

A large number of estimation methods for the transformation functions have been reported starting in the late 1980s. In the late 1980's and 90's, simple techniques employing vector quantisation (VQ) with codebooks~\cite{abe1988voice} or segmental codebooks~\cite{arslan1999speaker} of paired source-target frame vectors were proposed to represent the transformation functions. However, these VQ methods introduced frame-to-frame discontinuity problems.

In the late 1990's and 2000s,  \emph{joint density Gaussian mixture model} (JDGMM) based transformation methods~\cite{kain1998spectral,stylianou1998continuous} were proposed and have since then been actively improved by many researchers~\cite{toda2007voice,kobayashi2014statistical}. This method still remains popular even now. Although this method achieves smooth feature transformations using a locally linear transformation, this method also has several critical problems such as over-smoothing~\cite{popa2012local,chen2003voice,hwang2012study} and over-fitting~\cite{helander2010voice,pilkington2011gaussian} which leads to muffled quality of speech and degraded speaker similarity.

Therefore, in the early 2010, several alternative linear transformation methods were developed. Examples are partial least square (PLS) regression~\cite{helander2010voice}, tensor representation~\cite{saito2011one}, a trajectory HMM~\cite{zen2011continuous}, mixture of factor analysers~\cite{wu2012mixture}, local linear transformation~\cite{popa2012local} or noisy channel models~\cite{saito2012statistical}.

In parallel to the linear-based approaches, there have been studies on non-linear transformation functions such as support vector regression~\cite{song2011voice}, kernel partial least square~\cite{helander2012voicekernel}, and conditional restricted Boltzmann machines~\cite{wu2013conditional}, neural networks~\cite{narendranath1995transformation,desai2009voice}, highway network~\cite{SAITO20172017EDL8034}, and RNN~\cite{nakashika2015voice,7178896}. Data-driven frequency warping techniques~\cite{sundermann2003vtln,erro2010voice,erro2013voice} have also been studied.

Recently, deep learning has changed the above standard procedures for voice conversion and we can see many different solutions now. For instance, variational auto-encoder or sequence-to-sequence neural networks enable us to build VC systems without using frame level alignment~\cite{Hsu2017,Miyoshi2017}.  It has also been showed that a cycle-consistent adversarial network called ``CycleGAN''~\cite{cyclegan-vc} is one possible solution for building VC systems without using a parallel corpus. Wavenet can also be used as a replacement for the purpose of generating speech waveforms from converted acoustic features~\cite{Kobayashi2017}.

The approaches to voice conversion considered above are usually applied to the transformation of spectral envelope features, though the conversion of prosodic features such as fundamental frequency~\cite{gillet2003transforming,wu2006voice,helander2007novel,wu2010text} and duration~\cite{wu2006voice,lolive2008pitch} has also been studied.

For more details and technical comparisons, please see results of Voice Conversion Challenges that compare the performance of VC systems built on a common database~\cite{toda2016voice, wester2016analysis}.

\subsubsection{Spoofing}

When applied to spoofing, the aim with voice conversion is to synthesise a new speech signal such that the extracted ASV features are close in some sense to the target speaker. Some of the first works relevant to text-independent ASV spoofing were reported in~\cite{perrot2005voice,matrouf2006effect}.  The work in~\cite{perrot2005voice} showed that baseline EER increased from 16\% to 26\% thanks to a voice conversion system which also converted prosodic aspects not modeled in typical ASV systems. This work targeted the conversion of spectral-slope parameters and showed that the baseline EER of 10\% increased to over 60\% when all impostor test samples were replaced with converted voices. Moreover, signals subjected to voice conversion did not exhibit any perceivable artefacts indicative of manipulation.

The work in~\cite{kinnunen2012vulnerability} investigated ASV vulnerabilities to voice conversion based on JDGMMs~\cite{kain1998spectral} which requires a parallel training corpus for both source and target speakers.  Even if the converted speech could be easily detectable by human listeners, experiments involving five different ASV systems showed their universal susceptibility to spoofing. The FAR of the most robust, JFA system increased from 3\% to over 17\%. Instead of vocoder-based waveform generation, unit selection approaches can be applied directly to feature vectors coming from the target speaker to synthesise converted speech~\cite{sundermann2006text}.  Since they use target speaker data directly, unit-selection approaches arguably pose a greater risk to ASV than statistical approaches~\cite{Wu2013-textConstraint}. In the ASVspoof 2015 challenge, we therefore had chosen these popular VC methods as spoofing methods.

Other work relevant to voice conversion includes attacks referred to as artificial signals. It was noted in~\cite{alegre2012vulnerability} that certain short intervals of converted speech yield extremely high scores or likelihoods. Such intervals are not representative of intelligible speech but they are nonetheless effective in overcoming typical ASV systems which lack any form of speech quality assessment. The work in~\cite{alegre2012vulnerability} showed that artificial signals optimised with a genetic algorithm provoke increases in the EER from 10\% to almost 80\% for a GMM-UBM system and from 5\% to almost 65\% for a factor analysis (FA) system.

\subsubsection{Countermeasures}

Here, we provide an overview of countermeasure methods developed for the VC attacks before the ASVspoof challenge began.

Some of the first works to detect converted voice draws on related work in synthetic speech detection~\cite{deleon11}. In~\cite{wu2012detecting,wu2012study}, cosine phase and modified group delay function (MGDF) based countermeasures were proposed. These are effective in detecting converted speech using vocoders based on minimum phase. In VC, it is, however, possible to use natural phase information extracted from a source speaker \cite{matrouf2006effect}. In this case, they are unlikely to detect converted voice.

Two approaches to artificial signal detection are reported in~\cite{alegre2012spoofing}.  Experimental work shows that supervector-based SVM classifiers are naturally robust to such attacks, and that all the spoofing attacks they used could be detected by using an utterance-level variability feature, which detected the absence of the natural and dynamic variabilities characteristic of genuine speech. A related approach to detect converted voice is proposed in~\cite{alegre2013conversion}. Probabilistic mappings between source and target speaker models are shown to typically yield converted speech with less short-term variability than genuine speech. Therefore, the thresholded, average pair-wise distance between consecutive feature vectors was used to detect converted voice with an EER of under 3\%.

Due to fact that majority of VC techniques operate at the short-term frame level, more sophisticated long-term features such as temporal magnitude and phase modulation feature can also detect converted speech~\cite{wu2013synthetic}. Another experiment reported in~\cite{alegreInterspeech2013} showed that local binary pattern analysis of sequences of acoustic vectors can also be used for successfully detecting frame-wise JDGMM-based converted voice. However, it is unclear whether these features are effective in detecting recent VC systems that consider long-term dependency such as recurrent or autoregressive neural network models.

After the ASVspoof challenges took place, new countermeasures that works for both speech synthesis and voice conversion were proposed and evaluated. See the next section for a detailed review of the recently developed countermeasures.

\section{Summary of the spoofing challenges}

A number of independent studies confirm the vulnerability of ASV technology to spoofed voice created using voice conversion, speech synthesis, and playback~\cite{Wu2015survey}. Early studies on speaker anti-spoofing were mostly conducted on in-house speech corpora created using a limited number of spoofing attacks. The development of countermeasures using only a small number of spoofing
attacks may not offer the generalisation ability in the presence of different or unseen attacks. There was a lack of publicly available corpora and evaluation protocol to help with comparing the results obtained by different researchers.

The \index{ASVspoof}\footnote{\url{http://www.asvspoof.org/}} initiative aims to overcome this bottleneck by making available standard speech corpora consisting of a large number of spoofing attacks, evaluation protocols, and metrics to support a common evaluation and the benchmarking of different systems. The speech corpora were initially distributed by organising an evaluation challenge. In order to make the challenge simple and to maximise participation, the ASVspoof challenges so far involved only the detection
of spoofed speech; in effect, to determine whether a speech sample is genuine or spoofed. A training set and development set consisting of several spoofing attacks were first shared with the challenge participants to help them develop and tune their anti-spoofing algorithm. Next, the evaluation set without any label indicating genuine or spoofed speech was distributed, and the organisers asked the participants to submit scores within a specific deadline. Participants were allowed to submit scores of multiple systems. One of these systems was designated as the primary submission. Spoofing detectors for all primary submissions were trained using only the training data in the challenge corpus. Finally, the organisers evaluated the scores for benchmarks and ranking. The evaluation keys were subsequently released to the challenge participants. The challenge results were discussed with the participants in a special session in INTERSPEECH conferences, which also involved sharing knowledge and receiving useful feedback. To promote further research and technological advancements, the datasets used in the challenge are made publicly available.

The ASVspoof challenges have been organised twice so far. The first was held in 2015 and the second in 2017. A summary of the speech corpora used in the two challenges are shown in Table~\ref{Table:DatasetSummary}. In both the challenges, EER metric was used to evaluate the performance of spoofing detector. The EER is computed by considering the scores of genuine files as positive scores and those of spoofed files as negative scores. A lower EER means more accurate spoofing countermeasures. In practice, the EER is estimated using a specific \emph{receiver operating characteristics convex hull} (ROCCH) technique with an open-source implementation\footnote{\url{https://sites.google.com/site/bosaristoolkit/}} originating from outside the ASVspoof consortium. In the following subsections, we briefly discuss the two challenges. For more interested readers,~\cite{wu2015asvspoof} contains details of the 2015 edition while~\cite{kinnunen2017asvspoof} discusses the results of the 2017 edition.

\begin{table}[t]
\renewcommand{\arraystretch}{1.2}
\caption{Summary of the datasets used in ASVspoof challenges.}
\label{Table:DatasetSummary}
\vspace{0.2mm}
\centerline{
\begin{footnotesize}
\begin{tabular}{|c|c|c|}
\hline
 & \textbf{ASVspoof 2015}~\cite{wu2015asvspoof} & \textbf{ASVspoof 2017}~\cite{kinnunen2017asvspoof}\\
\hline
Theme &Detection of artificially generated speech& Detection of replay speech\\
\hline
Speech format  & $F_s$ = 16 kHz, 16 bit PCM & $F_s$ = 16 kHz, 16 bit PCM\\
\hline
Natural speech&Recorded using high-quality microphone& Recorded using different smart phones\\
\hline
Spoofed speech&Created with seven VC&Collected `in the wild' by crowdsourcing \\
& and three SS methods&using different microphone and playback  \\
&&devices from diverse environments\\
\hline
Spoofing types & 5 / 5 / 10 & 3 / 10 / 57\\
in train/dev/eval &&\\
\hline
No of speakers & 25 / 35 / 46   & 10 / 8 / 24\\
in train/dev/eval &&\\
\hline
No of genuine speech & 3750 / 3497 / 9404 & 1508 / 760 / 1298\\
files in train/dev/eval&&\\
\hline
No of spoofed speech & 12625 / 49875 / 184000& 1508 / 950 / 12008\\
files in train/dev/eval&&\\
\hline
\end{tabular}
\end{footnotesize}
}
\end{table}

\subsection{ASVspoof 2015}
The first ASVspoof challenge involved detection of artificial speech created using a mixture of voice conversion and speech synthesis techniques~\cite{wu2015asvspoof}. The dataset was generated with ten different artificial speech generation algorithms. The \index{ASVspoof 2015} was based upon a larger collection spoofing and anti-spoofing (SAS) corpus (v1.0)~\cite{wu2015icasspsas} that consists of both natural and artificial speech. Natural speech was recorded from 106 human speakers using a high-quality microphone and without significant channel or background noise effects. In a speaker disjoint manner, the full database was divided into three subsets called the training, development, and evaluation set. Five of the attacks (S1-S5), named as \emph{known attacks}, were used in the training and development set. The other five attacks, S6-S10, called \emph{unknown attacks}, were used only in the evaluation set, along with the known attacks. Thus, this provides the possibility of assessing the generalisability of the spoofing detectors. The detailed evaluation plan is available in~\cite{wu2014asvspoof}, describing the speech corpora and challenge rules.

Ten different spoofing attacks used in the ASVspoof 2015 are listed below:-
\begin{itemize}
  \item \textbf{S1}: a simplified frame selection (FS) based
voice conversion algorithm, in which the converted
speech is generated by selecting target speech frames.

  \item \textbf{S2}: the simplest voice conversion algorithm which
adjusts only the first mel-cepstral coefficient (C1) in order
to shift the slope of the source spectrum to the target.

  \item \textbf{S3}: a speech synthesis algorithm implemented with the
HMM based speech synthesis system (HTS3) using speaker adaptation techniques and
only 20 adaptation utterances.

  \item \textbf{S4}: the same algorithm as S3, but using 40 adaptation
utterances.

  \item \textbf{S5}: a voice conversion algorithm implemented with the
voice conversion toolkit and with the Festvox system\footnote{\url{http://www.festvox.org/}}.

  \item \textbf{S6}: a VC algorithm based on joint density Gaussian
mixture models (GMMs) and maximum likelihood parameter generation
considering global variance.

  \item \textbf{S7}: a VC algorithm similar to S6, but using line spectrum
pair (LSP) rather than mel-cepstral coefficients for
spectrum representation.

  \item \textbf{S8}: a tensor-based approach to VC, for which a
Japanese dataset was used to construct the speaker space.

  \item \textbf{S9}: a VC algorithm which uses kernel-based partial least
square (KPLS) to implement a non-linear transformation
function.

  \item \textbf{S10}: an SS algorithm implemented with the open-source
MARY text-to-tpeech system (MaryTTS)\footnote{\url{http://mary.dfki.de/}}.
\end{itemize}

More details of how the SAS corpus was generated can be found in~\cite{wu2015icasspsas}.

The organisers also confirmed the vulnerability to spoofing by conducting speaker verification experiments with this data and demonstrating considerable performance degradation in the presence of spoofing. With a state-of-the-art probabilistic linear discriminant analysis (PLDA) based ASV system, it is shown that in presence of spoofing, the average EER for ASV increases from 2.30\% to 36.00\% for male and 2.08\% to 39.53\% for female~\cite{wu2015asvspoof}. This motivates the development of the anti-spoofing algorithm.

For ASVspoof 2015, the challenge evaluation metric was the average EER. It is computed by calculating EERs for each attack and then taking average.  The
dataset was requested by 28 teams from 16 countries, 16 teams returned primary submissions by the deadline. A total of 27 additional submissions were also received. Anonymous results were subsequently returned to each team, who were then invited to submit their work to the ASVspoof special session for INTERSPEECH
2015.

\begin{table}[t]
\renewcommand{\arraystretch}{1.2}
\caption{Performance of top five systems in ASVspoof 2015 challenge (ranked according to the average \% EER for all attacks) with respective features and classifiers.}
\label{Table:SystemSummary2015}
\vspace{0.2mm}
\centerline{
\begin{footnotesize}
\begin{tabular}{|c|c|c|c|c|}
\hline
System & \multicolumn{3}{|c|}{Avg. EER for} & System\\
Identifier & known & unknown & all  &Description\\
\hline
A~\cite{patel2015combining}  & 0.408&2.013&1.211& \emph{Features:} mel-frequency cepstral coefficients (MFCC),\\
&&&& Cochlear filter cepstral coefficients plus instantaneous frequency (CFCCIF).\\
&&&& \emph{Classifier:} GMM.\\
\hline
B~\cite{Novoselov2015ASVspoof}  & 0.008&3.922&1.965& \emph{Features:} MFCC, MFPC,\\
&&&& cosine-phase principal coefficients (CosPhasePCs). \\
&&&& \emph{Classifier:} Support vector machine (SVM) with i-vectors.\\
\hline
C~\cite{chen2015robust}  & 0.058&4.998&2.528&\emph{Feature:} DNN-based with filterbank output and their deltas as input.\\
&&&& \emph{Classifier:} Mahalanobis distance on s-vectors.\\
\hline
D~\cite{xiao2015spoofing}  & 0.003&5.231&2.617& \emph{Features:} log magnitude spectrum (LMS),\\
&&&& residual log magnitude spectrum (RLMS), group delay (GD),\\
&&&& modified group delay (MGD), instantaneous frequency derivative (IF),\\
&&&& baseband phase difference (BPD), and pitch synchronous phase (PSP).\\
&&&& \emph{Classifier:} Multilayer perceptron (MLP).\\
\hline
E~\cite{alam2015development}  & 0.041&5.347&2.694& \emph{Features:}  MFCC, product spectrum MFCC (PS-MFCC),\\
&&&& MGD with and without energy, weighted linear prediction group delay\\
&&&&  cepstral coefficients (WLP-GDCCs), and MFCC\\
&&&&  cosine-normalised phase-based cepstral coefficients (MFCC-CNPCCs).\\
&&&& \emph{Classifier:} GMM.\\
\hline
\end{tabular}
\end{footnotesize}
}
\end{table}

Table~\ref{Table:SystemSummary2015} shows the performance of the top five systems in the ASVspoof 2015 challenge. The best performing system~\cite{patel2015combining} uses a combination of\emph{ mel cesptral} and \emph{cochlear filter cepstral coefficients plus instantaneous frequency} features with GMM back-end. In most cases, the participants have used fusion of multiple feature based systems to get better recognition accuracy. Variants of cepstral features computed from the magnitude and phase of short-term speech are widely used for the detection of spoofing attacks. As a back-end, GMM was found to outperform more advanced classifiers like i-vectors, possibly due to the use of short segments of high-quality speech not requiring treatment for channel compensation and background noise reduction. All the systems submitted in the challenge are reviewed in more detail~\cite{JSTSPOverview}.

\subsection{ASVspoof 2017}


The \index{ASVspoof 2017} is the second automatic speaker verification antispoofing and countermeasures challenge. Unlike the 2015 edition that used very high-quality speech material, the 2017 edition aims to assess  spoofing  attack  detection with ''out in the wild'' conditions. It focuses exclusively on replay attacks. The corpus originates from the recent \textit{text-dependent} \emph{RedDots} corpus\footnote{\url{https://sites.google.com/site/thereddotsproject/}}, whose purpose was to collect speech data over mobile devices, in the form of smartphones and tablet computers, by volunteers from across the globe.

The replayed version of the original \emph{RedDots} corpus was collected through a crowdsourcing exercise using various replay configurations consisting of varied devices, loudspeakers, and recording devices, under a variety of different environments across four European countries within the EU Horizon 2020-funded OCTAVE project\footnote{\url{https://www.octave-project.eu/}}, (see~\cite{kinnunen2017asvspoof}). Instead of covert recording, we made a ``short-cut'' and took the digital copy of the target speakers' voice to create the playback versions.
The collected corpus is divided into three subsets: for training, development, and evaluation.
Details of each are presented in Table~\ref{Table:DatasetSummary}.
All three subsets are disjoint in terms of speakers and data collection sites.
The training and development subsets were collected at three different sites. The evaluation subset was collected at the same three sites and also included data from two new sites. Data from the same site include different recordings and replaying devices and from different acoustic environments. The evaluation subset contains data collected from 161 replay sessions in 62 unique replay configurations\footnote{A \textbf{replay configuration} refers to a unique combination of room, replay device and recording device while a \textbf{session} refers to a set of source files, which share the same replay configuration.}. More details regarding replay configurations can be found in~\cite{kinnunen2017asvspoof,delgado2018asvspoof}.


The primary evaluation metric is ``pooled'' EER. In contrast to the ASVspoof 2015 challenge, the EER is computed from scores pooled across all the trial segments rather than condition averaging.
A baseline\footnote{See \textit{Appendix A.2. Software packages}} system based on common GMM back-end classifier with constant Q cepstral coefficient (CQCC)~\cite{Todisco16CQCC,Todisco2017CSL} features was provided to the participants. This configuration is chosen as baseline as it has shown best recognition performance on ASVspoof 2015. The baseline is trained using either combined training and development data (B01) or training data (B02) alone. The baseline system does not involve any kind of optimisation or tuning with respect to~\cite{Todisco16CQCC}. The dataset was requested by 113 teams, of which 49 returned primary submissions by the deadline. The results of the challenge were disseminated at a special session consisting of two slots at INTERSPEECH 2017.

\begin{figure*}[!t]
\centerline{\includegraphics[width=\linewidth]
{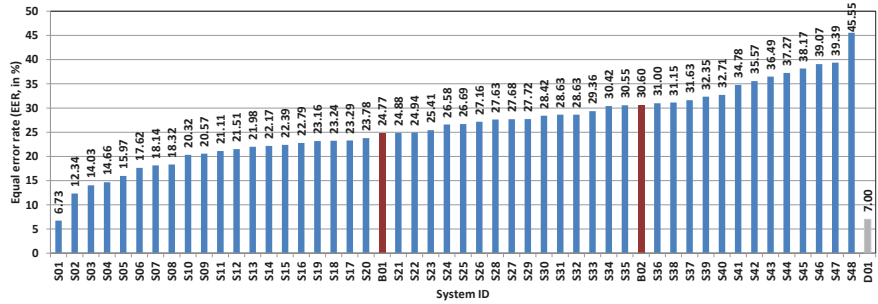}}
\caption{Performance of the two baseline systems (B01 and B02) and the 49 primary systems (S01---S48 in addition to late submission D01) for the ASVspoof 2017 challenge.  Results are in terms of the replay/non-replay EER (\%).}
\label{fig:challengeResults}
\end{figure*}

Most of the systems are based on standard spectral features, such as CQCCs, MFCCs, and \emph{perceptual linear prediction} (PLP). As a back-end, in addition to the classical GMM to model the replay and non-replay classes, it has also exploited the power of deep classifiers, such as \emph{convolutional neural network} (CNN) or \emph{recurrent neural network} (RNN). A fusion of multiple features and classifiers is also widely adopted by the participants. A summary of the top-10 primary systems is provided in Table~\ref{tab:participant-summary}. Results in terms of EER of the 49 primary systems and the baseline B01 and B02 are shown in Figure~\ref{fig:challengeResults}.

\begin{table*}[!t]
\renewcommand{\arraystretch}{1.4}
\caption{Summary of top 10 primary submissions to ASVspoof 2017. Systems' IDs are the same received by participants in the evaluation. The column `Training' refers to the part of data used for training: train (T) and/or development (D).}
\label{tab:participant-summary}
\begin{scriptsize}

\begin{tabular}{|p{1.2cm}|p{2cm}|p{1cm}|p{2.5cm}|p{.8cm}|p{.8cm}|p{1cm}|p{1.5cm}|}
\hline
ID	&		Features	&	Post-proc.	&	Classifiers	&	Fusion	&	\#Subs. & Training & Performances on eval subset (EER\%)\\
\hline\hline
S01~\cite{lavrentyeva2017audio}	&		Log-power Spectrum, LPCC	&	MVN	&	CNN, GMM, TV, RNN	&	Score 	&	3	&	T	& 6.73\\
S02~\cite{S02} &		CQCC, MFCC, PLP	&	WMVN	&	GMM-UBM, TV-PLDA, GSV-SVM, GSV-GBDT, GSV-RF	&	Score	&	--	&	T	& 12.34\\
S03&		MFCC, IMFCC, RFCC, LFCC, PLP, CQCC, SCMC, SSFC	&	--	&	GMM, FF-ANN	&	Score 	&	18	&	T+D	& 14.03\\
S04&		RFCC, MFCC, IMFCC, LFCC, SSFC, SCMC	&	--	&	GMM	&	Score 	&	12	&	T+D	& 14.66\\
S05~\cite{S05} &		Linear filterbank feature	&	MN	&	GMM, CT-DNN	&	Score 	&	2	&	T	& 15.97\\
S06 &		CQCC, IMFCC, SCMC, Phrase one-hot encoding	&	MN	&	GMM	&	Score 	&	4	&	T+D	& 17.62\\
S07 &		HPCC, CQCC	&	MVN	&	GMM, CNN, SVM	&	Score 	&	2	&	T+D	& 18.14\\
S08~\cite{S08} &		IFCC, CFCCIF, Prosody	&	--	&	GMM	&	Score 	&	3	&	T	& 18.32\\
S09	&		SFFCC	&	No	&	GMM	&	None	&	1	&	T & 20.57	\\
S10~\cite{S10} 	&		CQCC	&	--	&	ResNet	&	None	&	1	&	T &	20.32 \\
\hline

\end{tabular}
\end{scriptsize}
\end{table*}

\section{Advances in front-end features}

The selection of appropriate features for a given classification problem is an important task. Even if the classic boundary to think between a feature extractor (front-end) and a classifier (back-end) as separate components is getting increasingly blurred with the use of end-to-end deep learning and other similar techniques, research on the `early' components in a pipeline remains important. In the context of anti-spoofing for ASV, this allows the utilisation of one's domain knowledge to guide the design of new discriminative features. For instance, earlier experience suggests that lack of spectral~\cite{wu2012detecting} and temporal~\cite{wu2013synthetic} detail is characteristic of synthetic or voice-coded (vocoded) speech, and that low-quality replayed signals tend to experience loss of spectral details~\cite{wu2014study}. These initial findings sparked further research into developing advanced front-end features with improved robustness, generalisation across datasets, and other desideratum. As a matter of fact, in contrast to classic ASV (without spoofing attacks) where the most significant advancements have been in the back-end modelling~\cite{hansen2015speaker}, in ASV anti-spoofing, the features seem to make the difference. In this section, we take a brief look at a few such methods emerging from the ASVspoof evaluations. The list is by no means exhaustive and the interested reader is referred to~\cite{JSTSPOverview} for further discussion.

\subsection{Front-ends for detection of voice conversion and speech synthesis spoofing}


The front-ends described below have been shown to provide good performance on the ASVspoof 2015 database of spoofing attacks based on voice conversion and speech synthesis. The first front-end was used in the ASVspoof 2015 challenge, while the rest were proposed later after the evaluation.

\textbf{Cochlear filter cepstral coefficients with instantaneous frequency (CFCCIF).} These features were introduced in~\cite{patel2015combining} and successfully used as part of the top-ranked system in the ASVspoof 2015 evaluation. They combine cochlear filter cepstral coefficients (CFCC), proposed in~\cite{li2009auditory}, with instantaneous frequency~\cite{quatieri02}. CFCC are based on wavelet transform-like auditory transform and on some mechanisms of the cochlea of the human ear, such as hair cells and nerve spike density. To compute CFCC with instantaneous frequency (CFCCIF), the output of the nerve spike density envelope is multiplied by the instantaneous frequency, followed by the derivative operation and logarithm non-linearity. Finally, the discrete cosine transform (DCT) is applied to decorrelate the features and obtain a set of cepstral coefficients.

\textbf{Linear frequency cepstral coefficients (LFCC).}
LFCCs are very similar to the widely used mel-frequency cepstral coefficients (MFCCs)~\cite{DavisMermelstein1980}, though the filters are placed in equal sizes for linear scale. This front-end is widely used in speaker recognition and has been shown to perform well in spoofing detection~\cite{Sahidullah15Features}. This technique performs a windowing on the signal, computes the magnitude spectrum using the short-time Fourier transform (STFT), followed by logarithm non-linearity and the application of a filterbank of linearly-spaced $N$ triangular filters to obtain a set of $N$ log-density values. Finally, the DCT is applied to obtain a set of cepstral coefficients.

\begin{figure}[h]
\includegraphics[width=\columnwidth]{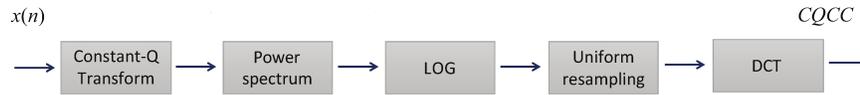}
\caption{Block diagram of CQCC feature extraction process.}
\label{fig:cqcc}
\end{figure}

\textbf{Constant Q cepstral coefficients (CQCC)}. This feature was proposed in \cite{Todisco16CQCC,Todisco2017CSL} for spoofing detection and it is based on the constant Q transform (CQT)~\cite{Brown91}. The CQT is an alternative time-frequency analysis tool to the STFT that provides variable time and frequency resolution. It provides greater frequency resolution at lower frequencies but greater time resolution at higher frequencies. Figure~\ref{fig:cqcc} illustrates the \index{CQCC} extraction process. The CQT spectrum is obtained, followed by logarithm non-linearity and by a linearisation of the CQT geometric scale. Finally, cepstral coefficients are obtained though the DCT.

As an alternative to CQCC, infinite impulse response constant-Q transform cepstrum (ICQC) features~\cite{Alam2017spoofing} use the infinite impulse response - constant Q transform~\cite{Cancela2009efficient}, an efficient constant Q transform based on the IIR filtering of the fast Fourier transform (FFT) spectrum. It delivers multiresolution time-frequency analysis in a linear scale spectrum which is ready to be coupled with traditional cepstral analysis. The IIR-CQT spectrum is followed by the logarithm and decorrelation, either through the DCT or principal component analysis.

\textbf{Deep features for spoofing detection.} All of the above three features sets are hand-crafted and consists of a fixed sequence of standard digital signal processing operations. An alternative approach, seeing increased popularity across different machine learning problems, is to learn the feature extractor from a given data by using deep learning techniques~\cite{bengio2009learning,goodfellow2016deep}. In speech-related
applications, these features are widely employed for improving recognition accuracy~\cite{Tian2015Bottleneck,Richardson2015deep,Hinton2012deep}. The work in~\cite{Alam2016Odyssey} uses deep neural network to generate bottleneck features for spoofing detection; that is, the activations of a hidden layer with a relatively small number of nodes compared to the size of other layers. The study in~\cite{Qian2016Deep} investigates various features based on deep learning techniques. Different feed-forward DNNs are used to obtain frame-level deep features. Input acoustic features consisting of filterbank outputs with their first derivatives are used to train the network to discriminate between the natural and spoofed speech classes, and output of hidden layers are taken as deep features which are then averaged to obtain an utterance-level descriptor. RNNs are also proposed to estimate utterance-level features from input sequences of acoustic features. In another recent work~\cite{Yu2017DNNCeps}, the authors have investigated deep features based on filterbank trained with the natural and artificial speech data. A feed forward neural network architecture called here as filterbank neural network (FBNN) is used here that includes a
linear hidden layer, a sigmoid hidden layer and a softmax output layer. The number of nodes in the output is six; and of them, five are for the number of spoofed classes in the training set, and the remaining one is for natural speech. The filterbanks are learned using the stochastic gradient descent algorithm. The cepstral features extracted using these DNN-based features are shown to be better than the hand-crafted cepstral coefficients.

\textbf{Scattering cepstral coefficients.} This feature for spoofing detection was proposed  in~\cite{Sriskandaraja2017frontend}. It relies upon \emph{scattering spectral decomposition}~\cite{anden2014deep,Mallat2012group}. This transform is a hierarchical spectral decomposition of a signal based on wavelet filter-banks (constant Q filters), modulus operator, and averaging. Each level of decomposition processes the input signal (either the input signal for the first level of decomposition, or the output of a previous level of decomposition) through the wavelet filterbank and takes the absolute value of filter outputs, producing a scalogram. The scattering coefficients at a certain level are estimated by windowing the scalogram signals and computing the average value within these windows. A two-level scattering decomposition has been shown to be effective for spoofing detection~\cite{Sriskandaraja2017frontend}. The final feature vector is computed by taking the DCT of the vector obtained by concatenating the logarithms of the scattering coefficients from all levels and retaining the first a few coefficients. The ``interesting'' thing about scattering transform is its stability to small signal deformation and more details of the temporal envelopes than MFCCs~\cite{anden2014deep,Sriskandaraja2017frontend}.

\textbf{Fundamental frequency variation features.} The prosodic features are not as successful as cepstral features in detecting artificial speech on ASVspoof 2015, though some earlier results on PAs indicate that pitch contours are useful for such tasks~\cite{Wu2015survey}. In a recent work~\cite{pal2018synthetic}, the author use fundamental frequency variation (FFV) for this. The FFV captures pitch variation at the frame-level and provides
complementary information on cepstral features~\cite{laskowski2008fundamental}. The combined system gives a very promising performance for both known and unknown conditions on ASVspoof evaluation data.

\textbf{Phase-based features.} The phase-based features are also successfully used in PAD systems for ASVspoof 2015. For example, relative phase shift (RPS) and modified group delay (MGD) based features are explored in~\cite{saratxaga2016synthetic}. The authors in~\cite{wang2017spoofing} have investigated relative phase information (RPI) features. Though the performances on seen attacks are promising with these phase-based features, the performances noticeably degrade for unseen attacks, particularly for S10.

\textbf{General observations regarding front-ends for artificial speech detection.} Beyond the feature extraction method used, there are two general findings common to any front end~\cite{Sahidullah15Features,patel2015combining,Todisco2017CSL,Alam2017spoofing}. The first refers to the use of dynamic coefficients. The first and second derivatives of the static coefficients, also known as velocity and acceleration coefficients, respectively are found important to achieve good spoofing detection performance. In some cases, the use of only dynamic features is superior to the use of static plus dynamic coefficients~\cite{Sahidullah15Features}. This is not entirely surprising, since voice conversion and speech synthesis techniques may fail to model the dynamic properties of the speech signals, introducing artefacts that help the discrimination of spoofed signals. The second finding refers to the use of speech activity detection. In experiments with ASVspoof 2015 corpus, it appears that the silence regions also contain useful information for discriminating between natural and synthetic speech. Thus, retaining non-speech frames turns out to be a better choice for this corpus~\cite{Sahidullah15Features}. This is likely due to the fact that non-speech regions are usually replaced with noise during the voice conversion or speech synthesis operation. However, this could be a database-dependent observation, thus detailed investigations are required.

\subsection{Front-ends for replay attack detection}
The following front-ends have been proposed for the task of replay spoofing detection, and evaluated in replayed speech databases such as the BTAS 2016 and ASVspoof 2017. Many standard front-ends, such as MFCC, LFCC, and PLP, have been combined to improve the performance of replay attack detection. Other front-ends proposed for synthetic and converted speech detection (CFCCIF, CQCC) have been successfully used for the replay detection task. In general, and in opposition to the trend for synthetic and converted speech detection, the use of static coefficients has been shown to be crucial for achieving good performance. This may be explained by the nature of the replayed speech detection task, where detecting changes in the channel captured by static coefficients helps with the discrimination of natural and replayed speech. Two additional front-ends are described next.

\textbf{Inverted mel frequency cepstral coefficients (IMFCC).} This front-end is relatively simple and similar to the standard MFCC. The only difference is that the filterbank follows an inverted mel scale; that is, it provides an increasing frequency resolution (narrower filters) when frequency increases, and a decreased frequency resolution (wider filters) for decreasing frequency, unlike the mel scale~\cite{chakroborty2009improved}. This front-end was used as part of the top-ranked system of the Biometrics: Theory, Applications, and Systems (BTAS) 2016 speaker antispoofing competition~\cite{Korshunov2016Overview}.

\textbf{Features based on convolutional neural networks.} In the recent ASVspoof 2017 challenge, the use of deep learning frameworks for feature learning was proven to be key in achieving good replay detection performance. In particular, convolutional neural networks have been successfully used to learn high-level utterance-level features which can later be classified with simple classifiers. As part of the top-ranked system~\cite{lavrentyeva2017audio} in the ASVspoof 2017 challenge, a light convolutional neural network architecture~\cite{Wu2015light} is fed with truncated normalised FFT spectrograms (to force fixed data dimensions). The network consists of a set of convolutional layers, followed by a fully-connected layer. The last layer contains two outputs with softmax activation corresponding to the two classes. All layers use the max-feature-map activation function~\cite{Wu2015light}, which acts as a feature selector and reduces the number of feature maps by half on each layer. The network is then trained to discriminate between the natural and spoofed speech classes. Once the network is trained, it is used to extract a high-level feature vector which is the output of the fully connected layer. All the test utterances are processed to obtain high-level representations, which are later classified with an external classifier.

\textbf{Other hand-crafted features.} Many other features have also been used for replayed speech detection in the context of the ASVspoof 2017 database. Even if the performances of single systems using such features are not always high, they are shown to be complementary when fused at the score level~\cite{Goncalves2017generalization}, similar to conventional ASV research outside of the spoofing detection. These features include MFCC, IMFCC, rectangular filter cepstral coefficients (RFCCs), PLP, CQCC, spectral centroid magnitude coefficients (SCMC), subband spectral flux coefficient (SSFC), and variable length Teager energy operator energy separation algorithm-instantaneous frequency cosine coefficients (VESA-IFCC). Though, of course, one usually then has to further train the fusion system, which makes the system more involved concerning practical applications.

\section{Advances in back-end classifiers}

In the natural vs. spoof classification problem, two main families of approaches have been adopted, namely generative and discriminative. Generative approaches include those of GMM-based classifiers and i-vector representations combined with support vector machines (SVMs). As for discriminative approaches, deep learning based techniques have become more popular. Finally, new deep learning end-to-end solutions are emerging. Such techniques perform the typical pipeline entirely through deep learning, from feature representation learning and extraction to the final classification. While including such approaches into the traditional classifiers category may not be the most precise, they are included in this classifiers section for simplicity.

\subsection{Generative approaches}

\textbf{Gaussian mixture model (GMM) classifiers.}
Considering two classes, namely natural and spoofed speech, one GMM can be learned for each class using appropriate training data. In the classification stage, an input utterance is processed to obtain its likelihoods with respect to the natural and spoofed models. The resulting classification score is the log-likelihood ratio between the two competing hypotheses; in effect, those of the input utterance belonging to the natural and to the spoofed classes. A high score supports the former hypothesis, while a low score supports the latter. Finally, given a test utterance, classification can be performed by thresholding the obtained score. If the score is above the threshold, the test utterance is classified as natural, and otherwise, it is classified as spoof. Many proposed anti-spoofing systems use GMM classifiers~\cite{Sahidullah15Features,patel2015combining,Todisco16CQCC,Paul16features,Alam2016Odyssey,Sriskandaraja2017frontend,Alam2017spoofing}.

\textbf{I-vector.}
The state-of-the-art \index{i-vector} paradigm for speaker verification~\cite{Dehak2011frontendFA} has been explored for spoofing detection~\cite{khoury2014introducing,Sizov2015joint}. Typically, an i-vector is extracted from an entire speech utterance and used as a low-dimensional, high-level feature which is later classified by means of a binary classifier, commonly cosine distance measure or support vector machine (SVM). Different amplitude- and phase-based frontends~\cite{Novoselov2015ASVspoof,lavrentyeva2017audio} can be employed for the estimation of i-vectors. A recent work shows that data selection for i-vector extractor training (also known as \textbf{T} matrix) is an important factor for achieving completive recognition accuracy~\cite{hanilcci2018data}.

\subsection{Discriminative approaches}

\textbf{DNN classifiers.}
Deep learning based classifiers have been explored for use in the task of natural and spoofed speech discrimination. In~\cite{Tian16FeatPerspective,Alam2016Odyssey}, several front-ends are evaluated with neural network classifier consisting of several hidden layers with sigmoid nodes and softmax output, which is used to calculate utterance posteriors. However, the implementation detail of the DNNs - such the number of nodes, the cost function, the optimization algorithm and the activation functions - is not precisely mentioned in those work and the lack of this very relevant information make it difficult to reproduce the results.

In a recent work~\cite{Yu2018IEEENNLS}, a five-layer DNN spoofing detection system is investigated for ASVspoof 2015 which uses a novel scoring method, termed in the paper as \emph{human log-likelihoods} (HLLs). Each of the hidden layers has 2048 nodes with a sigmoid activation function. The network has six softmax output layers. The DNN is implemented using a computational network toolkit\footnote{\url{https://github.com/Microsoft/CNTK}} and trained with stochastic gradient descent methods with dynamics information of acoustic features, such as spectrum-based cepstral coefficients (SBCC) and CQCC as input. The cross entropy function is selected as the cost function and the maximum training epoch is chosen as 120. The mini-batch size is set to 128. The proposed method shows considerable PAD detection performance. The author obtain an EER for S10 of 0.255\% and average EER for all attacks of 0.045\% when used with CQCC acoustic features. These are the best reported performance in ASVspoof 2015 so far.

\textbf{DNN-based end-to-end approaches.} End-to-end systems aim to perform all the stages of a typical spoofing detection pipeline, from feature extraction to classification, by learning the network parameters involved in the process as a whole. The advantage of such approaches is that they do not explicitly require prior knowledge of the spoofing attacks as required for the development of acoustic features. Instead, the parameters are learned and optimised from the training data. In \cite{Dinkel2017end}, a convolutional long short-term memory (LSTM) deep neural network (CLDNN)~\cite{sainath2015learning} is used as an end-to-end solution for spoofing detection. This model receives input in the form of a sequence of raw speech frames and outputs a likelihood for the whole sequence. The CLDNN performs time-frequency convolution through CNN to reduce spectral variance, long-term temporal modelling by using a LSTM, and classification using a DNN. Therefore, it is a entirely an end-to-end solution which does not rely on any external feature representation. The works in~\cite{Zhang2017investigation,lavrentyeva2017audio} propose other end-to-end solutions by combining convolutional and recurrent layers, where the first act as a feature extractor and the second models the long-term dependencies and acts as a classifier. Unlike the work in ~\cite{Dinkel2017end}, the input data is the FFT spectrogram of the speech utterance and not the raw speech signal. In~\cite{Muckenhirn2017IJCB}, the authors have investigated CNN-based end-to-end system for PAD where the raw speech is used to jointly learn the feature extractor and classifier. Score-level combination of this CNN system with standard long-term spectral statistics based system shows considerable overall improvement.



\section{Other PAD approaches}
While most of the studies in voice PAD detection research focus on algorithmic improvements for discriminating natural and artificial speech signals, some recent studies have explored utilising additional information collected using special additional hardware to protect ASV system from presentation attacks~\cite{chen2017you,shiota2015voice,shiota2016voice,sahidullah2018robust}. Since an intruder can easily collect voice samples for the target speakers using covert recording; the idea there is to detect and recognise supplementary information related to the speech production process. Moreover, by its nature, that supplementary information is difficult, if not impossible, to mimic using spoofing methods in the practical scenario. These PAD techniques have shown excellent recognition accuracy in the spoofed condition, at the cost of additional setup in the data acquisition step.

The work presented in~\cite{shiota2015voice,shiota2016voice} utilises the phenomenon of \index{\emph{pop noise}}, which is a distortion in human breath when it reaches a microphone~\cite{elko2007electronic}. During natural speech production, the interactions between the airflow and the vocal cavities may result in a sort of plosive burst, commonly know as pop noise, which can be captured via a microphone. In the context of professional audio and music production, pop noise is unwanted and is eliminated during the recording or mastering process. In the context of ASV, however, it can help in the process of PAD. The basic principle is that a replay sound from a loudspeaker does not involve the turbulent airflow generating the pop noise as in the natural speech. The authors in~\cite{shiota2015voice,shiota2016voice} have developed a pop noise detector which eventually distinguishes natural speech from playback recording as well as synthetic speech generated using VC and SS methods. In experiments with 17 female speakers, a tandem detection system that combines both single- and double-channel pop noise detection gives the lowest ASV error rates in the PA condition.

The authors in~\cite{chen2017you} have introduced the use of a smartphone-based \emph{magnetometer} to detect voice presentation attack. The conventional loudspeakers, which are used for playback during access of the ASV systems, generate sound using acoustic transducer and generate a magnetic field. The idea, therefore, is to capture the use of loudspeaker by sensing the magnetic field which would be absent from human vocals. Experiments were conducted using playback from 25 different conventional loudspeakers, ranging from low-end to high-end and placed in different distances from the smartphone that contains the ASV system. A speech corpus of five speakers was collected for the ASV experiments executed using an open-source ASV toolkit, SPEAR\footnote{\url{https://www.idiap.ch/software/bob/docs/bob/bob.bio.spear/stable/index.html}}. Experiments were conducted with other datasets, using a similarly limited number of speakers. The authors demonstrated that the magnetic field based detection can be reliable for the detection of playback within 6-8~cm from the smartphone. They further developed a mechanism to detect the size of the sound source to prevent the use of small speakers, such as ear phones.

The authors in~\cite{zhang2016voicelive,zhang2017hearing} utilise certain acoustics concepts to prevent ASV systems from PAs. They first introduced a method~\cite{zhang2016voicelive} that estimates dynamic sound source position (articulation position within mouth) of some speech sounds using a small array using \emph{microelectromechanical systems} (MEMS) microphones embedded in mobile devices and compare it with loudspeakers, which have a flat sound source. In particular, the idea is to capture the dynamics of \emph{time-difference-of-arrival} (TDOA) in a sequence of speech sounds to the microphones of the smartphone. Such unique TDOA changes, which do not exist under replay conditions, are used for detecting replay attacks. The similarities between the TDOAs of test speech and user templates are measured using probability function under Gaussian assumption and correlation measure as well as their combinations. Experiments involving 12 speakers and three different types of smartphone demonstrate a low EER and high PAD accuracy. The proposed method is seen to remain robust despite the change of smartphones during the test and the displacements.

In ~\cite{zhang2017hearing}, the same research group has used the idea of the \emph{Doppler effect} to detect the replay attack. The idea here is to capture the \emph{articulatory gestures} of the speakers when they speak a pass-phrase. The smartphone acts as a Doppler radar and transmits a high frequency tone at 20 kHz from the built-in speaker and senses the reflections using the microphone during authentication process. The movement of the speaker's articulators during vocalisation creates a speaker-dependent Doppler frequency shift at around 20 kHz, which is stored along with the speech signal during the speaker-enrolment process. During a playback attack, the Doppler frequency shift will be different due to the lack of articulatory movements. Energy-based frequency features and frequency-based energy features are computed from a band of 19.8 kHz and 20.2 kHz. These features are used to discriminate between the natural and replayed voice; and the similarity scores are measured in terms of Pearson correlation coefficient. Experiments are conducted with a dataset of 21 speakers and using three different smartphones. The data also includes test speech for replay attack with different loudspeakers and for impersonation attack with four different impersonators. The proposed system was demonstrated to be effective in achieving low EER for both types of attacks. Similar to~\cite{zhang2016voicelive}, the proposed method indicated robustness to the phone placement.

\begin{figure}
\centering
  \includegraphics[width=\linewidth]{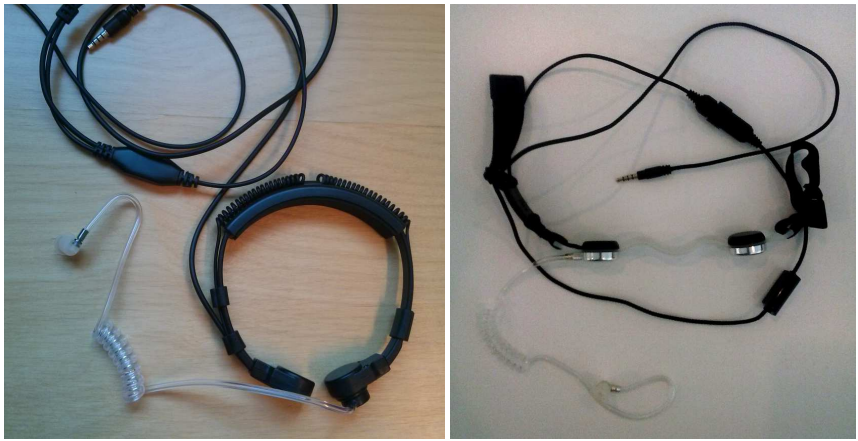}
  \caption{Throat-microphones used in~\cite{sahidullah2018robust} [Reprinted with permission from IEEE\/ACM Transactions on (T-ASL) Audio, Speech, and Language Processing].}
  \label{Fig:TM}
\end{figure}

The work in~\cite{sahidullah2018robust} introduces the use of a specific non-acoustic sensor, \emph{throat microphone} (TM), or laryngophone, to enhance the performance of the voice PAD system. An example of such microphones is shown in Fig.~\ref{Fig:TM}. The TM is used with a conventional acoustic microphone (AM) in a dual-channel framework for robust speaker recognition and PAD. Since this type of microphone is attached to the speaker's neck, it would be difficult for the attacker to obtain a covert recording of the target speaker's voice. Therefore, one possibility for the intruder is to use the stolen recording from an AM and to try to record it back using a TM for accessing the ASV system. A speech corpus of 38 speakers was collected for the ASV experiments. The dual-channel setup yielded considerable ASV for both licit and spoofed conditions. The performance is further improved when this ASV system is integrated with the dual-channel based PAD. The authors show zero FAR for replay imposters by decision fusion of ASV and PAD.

All of the above new PAD methods deviating from the ``mainstream'' of PAD research in ASV are reported to be reliable and useful in specific application scenarios for identifying presentation attacks. The methods are also fundamentally different and difficult to compare in the same settings. Since the authors focus on the methodological aspects, experiments are mostly conducted on a dataset of limited number of speakers. Extensive experiments with more subjects from diverse environmental conditions should be performed to assess their suitability for real-world deployment.

\section{Future directions of anti-spoofing research}
The research in ASV anti-spoofing is becoming popular and well-recognised in the speech processing and voice-biometric community. The state-of-the-art spoofing detector gives promising accuracy in the benchmarking of spoofing countermeasures. Further work is needed to address a number of specific issues regarding its practical use. A number of potential topics for consideration in further work are now discussed.

\begin{itemize}
  \item \textbf{Noise, reverberation and channel effect.} Recent studies indicate that spoofing countermeasures offer little resistance
to additive noise~\cite{Hanilci2016-noisy-spoof,Yu2016Multicondition}, reverberation~\cite{Tian16IS} and channel effect~\cite{Delgado2017BIOSIG} even though their performances on ``clean'' speech corpus are highly promising. The relative degradation of performance is actually much worse than the degradation of a typical ASV system under the similar mismatch condition. One reason could be that, at least until the ASVspoof 2017 evaluation, the methodology developed has been driven in clean, high-quality speech. In other words, the community might have developed its methods implicitly for laboratory testing. The commonly used speech enhancement algorithms also fail to reduce the mismatch due to environmental differences, though multi-condition training~\cite{Yu2016Multicondition} and more advanced training methods~\cite{Qian2017TASLP} have been found useful. The study presented in~\cite{Delgado2017BIOSIG} shows considerable degradation of PAD performance even in \emph{matched} acoustic conditions. The feature settings used for the original corpus gives lower accuracy when both training and test data are digitally processed with the telephone channel effect. These are probably because the spoofing artefacts themselves act as extrinsic variabilities which degrade the speech quality in some way. Since the task of spoofing detection is related to detecting those artefacts, the problem becomes more difficult in the presence of small external effects due to variation in environment and channel. These suggests further investigations need to be carried out for the development of robust spoofing countermeasures.

  \item \textbf{Generalisation of spoofing countermeasures.} The \index{generalisation} property of spoofing countermeasures for detecting new kinds of speech presentation attack is an important requirement for their application in the wild. Study explores that countermeasure methods trained with a class of spoofing attacks fail to generalise this for other classes of spoofing attack~\cite{Korshunov16IS,Goncalves2017generalization}. For example, PAD systems trained with VC and SS based spoofed speech give a very poor performance for playback detection~\cite{paul2017generalization}. The results of the first two ASVspoof challenges also reveal that detecting the converted speech created with an ``unknown'' method or the playback voice recording in a new replay session are difficult to detect. These clearly indicate the overfitting of PAD systems with available training data. Therefore, further investigation should be conducted to develop attack-independent universal spoofing detector. Other than the unknown attack issue, generalisation is also an important concern for cross-corpora evaluation of the PAD system~\cite{lorenzo2018can}. This specific topic is discussed in chapter 19 of this book.

  \item \textbf{Investigations with new spoofing methods.} The studies of converted spoof speech mostly focused on methods based on classical signal processing and machine learning techniques. Recent advancements in VC and SS research with deep learning technology show significant improvements in creating high quality synthetic speech~\cite{van2016wavenet}. The GAN~\cite{DBLP:conf/nips/GoodfellowPMXWOCB14} can be used to create (generator) spoofed voices with relevant feedback from the spoofing countermeasures (discriminator). Some preliminary studies demonstrate that the GAN-based approach can make speaker verification systems more vulnerable to presentation attacks~\cite{kreuk2018fooling,cai2018attacking}. More detailed investigations should be conducted on this direction for the development of countermeasure technology to guard against this type of advanced attack.

  \item \textbf{Joint operations of PAD and ASV.} The ultimate goal of developing PAD system is to protect the recogniser, the ASV system from imposters with spoofed speech. So far, the majority of the studies focused on the evaluation of standalone countermeasures. The integration of these two systems is not trivial number of reasons. First, standard linear output score fusion techniques, being extensively used to combine homogenous ASV system, are not appropriate since the ASV and its countermeasures are trained to solve two different tasks. Second, an imperfect PAD can increase the false alarm rate by rejecting genuine access trials~\cite{Sahidullah16jointCM-ASV}. Thirdly, and more fundamentally, it is not obvious whether improvements in standalone spoofing countermeasures should improve the overall system as a whole: a nearly perfect PAD system with close to zero EER may fail to protect ASV system in practice if not properly calibrated~\cite{muckenhirn2017long}. In a recent work~\cite{sarkar2017improving}, the authors propose a modification in a GMM-UBM based ASV system to make it suitable for both licit and spoofed conditions. The joint evaluation of PAD and ASV, as well as their combination techniques, certainly deserves further attention. Among other feedback received from the attendees of the ASVspoof 2017 special session organised during INTERSPEECH 2017, it was proposed that the authors of this chapter consider shifting the focus from standalone spoofing to more ASV-centric solutions in future. We tend to agree. In our recent work~\cite{kinnunen2018t}, we propose a new cost function for joint assessment of PAD and ASV system. In another work~\cite{todiscointegrated}, we propose a new fusion method for combining scores of countermeasures and recognisers. This work also explores speech features which can be used both for PAD and ASV.


\end{itemize}

\section{Conclusion}
This contribution provides an introduction to the different voice presentation attacks and their detection methods. It then reviews previous works with a focus on recent progress in assessing the performance of PAD systems. We have also briefly reviewed two recent ASVspoof challenges organised for the detection of voice PAs. This study includes discussion of recently developed features and the classifiers which are predominantly used in ASVspoof evaluations. We further include an extensive survey on alternative PAD methods. Apart from the conventional voice-based systems that use statistical properties of natural and spoofed speech for their discrimination, these recently developed methods utilise a separate hardware for the acquisition of other signals such as pop noise, throat signal, and extrasensory signals with smartphones for PAD. The current status of these non-mainstream approaches to PAD detection is somewhat similar to the status of the now more-or-less standard methods for artificial speech and replay PAD detection some three to four years ago: they are innovative and show promising results, but the pilot experiments have been carried out on relatively small and/or proprietary datasets, leaving an open question as to how scalable or generalisable these solutions are in practice. Nonetheless, in the long run and noting especially the rapid development of speech synthesis technology, it is likely that the quality of artificial/synthetic speech will eventually be indistinguishable from that of natural human speech. Such future spoofing attacks therefore could not be detected using the current mainstream techniques that focus on spectral or temporal details of the speech signal, but will require novel ideas that benefit from auxiliary information, rather than just the acoustic waveform.

In the past three years, the progress in voice PAD research has been accelerated by the development and free availability of speech corpus such as the ASVspoof series, SAS, BTAS 2016, AVSpoof. The work discussed several open challenges which show that this problem requires further attention to improving robustness due to mismatch condition, generalisation to new type of presentation attacks, and so on. Results from joint evaluations with integrated ASV system are also an important requirement for practical applications of PAD research. We think, however, that this extensive review will be of interest not only to those involved in voice PAD research but also to voice-biometrics researchers in general.

\section*{Appendix A. Action towards reproducible research}

\subsection*{A.1. Speech corpora}

\begin{enumerate}
\item Spoofing and Anti-Spoofing (SAS) database v1.0: This database presents the first version of a speaker verification spoofing and anti-spoofing database, named SAS corpus~\cite{wu2016anti}. The corpus includes nine spoofing techniques, two of which are speech synthesis, and seven are voice conversion.

Download link: \url{http://dx.doi.org/10.7488/ds/252}

\item ASVspoof 2015 database: This database has been used in the first Automatic Speaker Verification Spoofing and Countermeasures Challenge (ASVspoof 2015). Genuine speech is collected from 106 speakers (45 male, 61 female) and with no significant channel or background noise effects. Spoofed speech is generated from the genuine data using a number of different spoofing algorithms. The full dataset is partitioned into three subsets, the first for training, the second for development and the third for evaluation.

Download link: \url{http://dx.doi.org/10.7488/ds/298}

\item ASVspoof 2017 database: This database has been used in the Second Automatic Speaker Verification Spoofing and Countermeasuers Challenge: ASVspoof 2017. This database makes an extensive use of the recent text-dependent RedDots corpus, as well as a replayed version of the same data. It contains a large amount of speech data from 42 speakers collected from 179 replay sessions in 62 unique replay configurations.

Download link: \url{http://dx.doi.org/10.7488/ds/2313}

\end{enumerate}

\subsection*{A.2. Software packages}
\begin{enumerate}
\item Feature extraction techniques for anti-spoofing: This package contains the MATLAB implementation of different acoustic feature extraction schemes as evaluated in~\cite{Sahidullah15Features}.

Download link: \url{http://cs.joensuu.fi/~sahid/codes/AntiSpoofing_Features.zip}

\item Baseline spoofing detection package for ASVspoof 2017 corpus: This package contain the MATLAB implementations of two spoofing detectors employed as baseline in the official ASVspoof 2017 evaluation. They are based on constant Q cepstral coefficients (CQCC)~\cite{Todisco2017CSL} and Gaussian mixture model classifiers.

Download link: \url{http://audio.eurecom.fr/software/ASVspoof2017_baseline_countermeasures.zip}

\end{enumerate}

\ifthenelse{\equal{false}{\buildbook}}{
\printindex
\printglossary
\bibliographystyle{unsrt}
\bibliography{references}
}

\end{document}